\documentclass[journal=jacsat,manuscript=article]{achemso}
\setkeys{acs}{maxauthors=100} 
\usepackage[symbol]{footmisc}
\usepackage{multirow}
\usepackage{amsmath}
\usepackage{physics}
\usepackage[utf8]{inputenc}
\usepackage{booktabs}
\usepackage{float}
\usepackage{caption}
\usepackage{subcaption}
\usepackage{graphicx}
\usepackage{multicol}
\usepackage{siunitx}
\usepackage{soul}

\usepackage{tablefootnote}

\renewcommand{\ket}[1]{\vert #1 \rangle}

\renewcommand{\bra}[1]{\langle #1 \vert}

\newcommand{\Dbraket}[2]{\langle #1 \hspace{.10em} \vert \hspace{.10em}  #2 \rangle}

\newcommand{\Tbraket}[3]{\langle #1 \hspace{.10em} \vert \hspace{.10em} #2 \hspace{.10em} \vert \hspace{.10em} #3 \rangle}

\newcommand{\ketR}{\ket{\mathrm{R}}}

\newcommand{\simH}{\bar{H}}
\newcommand{\simP}{\bar{\mathcal{P}}}
\newcommand{\simS}{\bar{{S}}}
\renewcommand{\P}{\mathcal{P}}

\newcommand{\braR}{\bra{\mathrm{R}}}

\usepackage[dvipsnames]{xcolor}

\title{Excitation Energies from the Entanglement Coupled Cluster Model for Doublets}
\date{\today}
\author{Sarai Dery Folkestad}
\affiliation{Department of Chemistry, Norwegian University of Science and Technology, NTNU, 7491 Trondheim, Norway}
\email{sarai.d.folkestad@ntnu.no}
\author{Kristine Lauvstad Kruken}
\affiliation{Department of Chemistry, Norwegian University of Science and Technology, NTNU, 7491 Trondheim, Norway}
\author{Henrik Koch}
\affiliation{Department of Chemistry, Norwegian University of Science and Technology, NTNU, 7491 Trondheim, Norway}

\begin{document}
\maketitle
\begin{abstract}
We present excitation energies for molecular doublets from a spin-adapted formulation of coupled cluster singles and doubles theory. The entanglement coupled cluster approach represents an unconventional take on the notorious problem of spin adaptation for open-shell species. 
In this approach, the high-spin open-shell molecular system is coupled to non-interacting bath orbitals to form a total closed-shell system. In entanglement coupled cluster theory, many of the attractive features of the spin-adapted closed-shell coupled cluster is retained: an unambiguous definition of the cluster operator and a terminating Baker-Campbell-Hausdorff expansion. The result is a spin-adapted coupled cluster theory for open-shell species. 
The model produces excitation energies of a quality comparable to the closed-shell counterpart. Additionally, some ionized states that cannot be modeled accurately with the alternative equation-of-motion approach for ionized states, can be described with the entanglement coupled cluster singles and doubles model.
\end{abstract}

\section{Introduction}
The non-relativistic molecular electronic Hamiltonian commutes with the spin operators, and hence, the exact electronic wave function is also an eigenfunction of the $S^2$ and $S_z$ operators.\citep{helgaker2014} In electronic structure theory, a model is spin-adapted if it preserves these spin properties of the exact wave function. While enforcing the wave function to be an eigenfunction of the $S_z$ operator is trivial, satisfying the condition that it also is an eigenfunction of $S^2$ may be complicated for spin quantum number $S>0$, depending on the parametrization of the wave function. 
Developing a spin-adapted formulation of coupled cluster theory for open-shell species has turned out to be a major challenge. 

An alternative to spin-adapted coupled cluster theory, is the unrestricted, or spin-orbital formulation. Unrestricted coupled cluster theory is simple in both formulation and implementation. However, since the requirement that the wave function is an eigenfunction of the ${S}^2$ operator is lifted, the approach suffers from spin contamination. While the spin contamination of the unrestricted coupled cluster ground state\cite{stanton1994} is limited, even for systems where the UHF spin contamination is significant, spin contamination can be substantial in excited states. Furthermore,  magnetic properties depend explicitly on spin and are likely more sensitive to spin contamination than the energies. 

For closed-shell species, spin-adaptation of coupled cluster theory is straightforward. The cluster operator is defined in terms of singlet (spin-free) excitation operators, which preserves the spin properties of the reference determinant. Since all molecular orbitals are either doubly occupied or vacant, all terms in the cluster operator commute and the  Baker-Campbell-Hausdorff expansion of the similarity transformed Hamiltonian terminates after four nested commutators.  

A spin-adapted theory for high-spin open-shell systems can be formulated using the same strategy by constructing the cluster operator from singlet excitation operators relative to an open-shell reference determinant. However, for open-shell systems there are several complications with this strategy. 

First, there is freedom in the choice of the cluster operator, unlike in the closed-shell theory, and 
cluster operators have been suggested for open-shell systems
that generate either non-orthogonal \cite{janssen1991automated,herrmann2020generation,herrmann2022analysis} or orthogonal\citep{li1994automation, li1995spin, li1995unitary}  configurations.
While all these approaches yield a spin-adapted theory, the requirement of spin completeness, i.e., that the parametrization sufficiently spans the spin space for all included spatial configurations\citep{herrmann2020generation} requires the inclusion of spectator excitations of higher order than the excitation rank of the model. E.g., in coupled cluster singles and doubles (CCSD), triple and quadruple spectator excitations must be included in the cluster operator to obtain spin completeness.\cite{herrmann2020generation, herrmann2022correctly}

Secondly, there are non-commuting terms in the cluster operator, caused by the singly occupied orbitals of the reference determinant.  In the spin-free formulation, these orbitals appear both as hole and particle indices in the cluster operator. As a result, the spin-adapted open-shell equations become intractable, as the Baker-Campbell-Hausdorff expansion no longer terminates after four nested commutators. This means that 1) spin-adapted open-shell coupled cluster theory is derived and implemented in an automated manner through the use of symbolic computer programs, and 2) that the Baker-Campbell-Hausdorff expansion must be truncated at some order, or some other approximations must be introduced. This non-terminating feature of the open-shell coupled cluster equations has therefore motivated the development of models\citep{nooijen1996general, nooijen2001towards, datta2008compact, datta2013non, gunasekera2024multi} using a normal ordered exponential operator.\cite{lindgren1978coupled} 

The models mentioned so far obtain a spin-adapted formulation through the use of singlet (spin-free) excitation operators. 
A different approach was suggested by Heckert \textit{et al.},\citep{heckert2006towards} who formulated a spin-adapted theory where the cluster amplitudes are determined by solving simultaneously a non-redundant set of projected Schrödinger and spin equations. Heckert \textit{et al.} define the cluster operator in terms of excitations that generate configuration state functions, which are also used as a projection space when they solve the equations. Hence, as in the spin-adapted theories formulated in the molecular orbital basis, their amplitudes are spin-free, and their formulation in terms of configuration state functions ensures spin completeness.  While their approach requires the inclusion of excitations of rank three in the cluster operator in their spin-adapted CCSD model, their approach is fundamentally formulated in the spin-orbital basis and, therefore, there are no non-commuting terms in their cluster operator. 

The problem with spin contamination in unrestricted coupled cluster theory, and the major challenge evident in the formulation of spin-adapted theories, has motivated the development of models that aims to partially enforce the spin properties of the wave function. \cite{jayatilaka1993open, knowles1993coupled, neogrady1994spin, szalay1997spin}

 Coupled cluster theory, in the closed-shell spin-adapted formulation and in the spin-orbital formulation, has been an important tool for the calculation of excitation energies through the use of linear response\citep{koch1990coupled,pedersen1997coupled} theory or the equation-of-motion\citep{sekino1984linear,geertsen1989equation,bartlett2012coupled,krylov2008equation} approach. 
As is evident from the discussion above, significant efforts have gone into the formulation of a coupled cluster theory of the ground state wave function of high-spin open-shell systems. For excited states, on the other hand, few results have been published.

In addition to excitation energies, the equation-of-motion framework offers a way to describe ionized or electron attached wave functions (e.g., with $S = \frac{1}{2}$), even in a spin-adapted closed-shell implementation of the theory.  The approach was introduced by Stanton and Gauss\cite{stanton1999simple}, who included a non-interacting (ultra-diffuse) virtual (occupied) orbital in their calculation and ensured that an electron is excited into (out of) that orbital in the excited state calculation. With this approach, both the ground and excited states of an anionic or cationic system is available on equal footing.  This approach is simple to implement through projections during the diagonalization of the similarity transformed Hamiltonian. However, forcing an electron into, or out of, the non-interacting orbital reduces the parametric freedom in the EOM calculation and higher orders of coupled cluster theory is sometimes  needed to achieve accuracy comparable to that of standard valence excited states. 

Previously, we have presented a proof-of-concept implementation of the entanglement coupled cluster singles and doubles (ECCSD) model for the ground state of doublet systems.\citep{folkestad2023entanglement}
With entanglement coupled cluster theory, we describe high-spin open-shell systems in a spin-adapted manner. For doublet systems, a non-interacting orbital is mixed with the singly occupied molecular orbital of the system.  A closed-shell reference determinant for coupled cluster theory is constructed in this mixed orbital basis and through the use of a projection operator, we ensure that the molecule has the correct electron number and spin state.  
Here,
we present an $\mathcal{O}(N^6)$-scaling implementation of the entanglement coupled cluster singles and doubles model for both  ground and excited states of doublet molecular systems.

\section{Theory}
We consider a doublet molecular system, with molecular orbitals $\{\psi_p\}$ from a restricted open-shell Hartree--Fock calculation, and include in our calculation a non-interacting bath orbital $\psi_{\mathcal{B}}$. The non-relativistic molecular electronic Hamiltonian is given by
\begin{align}
\begin{split}
    H &= \sum_{pq} {h}_{pq} E_{pq} + \frac{1}{2}\sum_{pqrs}{g}_{pqrs}(E_{pq}E_{rs} - \delta_{qr}E_{ps})\\
    &+ {h}_{\mathcal{B}}E_{\mathcal{BB}} + \frac{1}{2}{g}_{\mathcal{B}}(E_{\mathcal{BB}}E_{\mathcal{BB}} - E_{\mathcal{BB}}), \label{eq:hamiltonian}
\end{split}
\end{align}
where 
\begin{align}
    E_{pq} = a^\dagger_{p\alpha}a_{q\alpha} + a^\dagger_{p\beta}a_{q\beta} = E_{pq}^{\alpha} + E_{pq}^{\beta}
\end{align}
is a singlet excitation operator, and where $h_{pq}$ and $g_{pqrs}$ are the one- and two-electron integrals (the latter in Mulliken ordering). 
The summations in Eq. \eqref{eq:hamiltonian} are restricted to the molecular orbitals, while ${h}_{\mathcal{B}}$ and ${g}_{\mathcal{B}}$ define the one- and two-electron interactions within the non-interacting orbital.

We mix the non-interacting orbital with the singly occupied molecular orbital (SOMO) of the molecular doublet $\psi_{\mathcal{S}}$, 
\begin{align}
\psi_I &= \psi_{\mathcal{S}}\cos\theta - \psi_{\mathcal{B}}\sin\theta\label{eq:occ_mixed}\\
\psi_A &=  \psi_{\mathcal{S}}\sin\theta +  \psi_{\mathcal{B}}\cos\theta,\label{eq:vir_mixed}
\end{align}
and thereby obtain a new (mixed) orbital basis, defined by the mixing angle $\theta$.
This basis includes the orbitals of the original basis, except for the orbitals $\psi_{\mathcal{S}}$ and $\psi_{\mathcal{B}}$, which have been replaced by $\psi_I$ and $\psi_A$.

From now on, we let $i, j, k$ denote the doubly occupied orbitals of the molecular doublet, $a, b, c$ denote the virtual orbitals of the molecular doublet, and $p, q, r, s$ denote orbitals, in general. The indices $I$ and $A$ are given the orbitals resulting from mixing $\psi_{\mathcal{S}}$ and $\psi_{\mathcal{B}}$.

In this mixed orbital basis, we construct a closed-shell reference according to
\begin{align}
\begin{split}
        \ketR &= {a}^\dagger_{I\alpha}{a}^\dagger_{I\beta}\prod_{i}{a}^\dagger_{i\alpha}{a}^\dagger_{i\beta}
    \ket{\text{vac}}
    \label{eq:reference_state}
\end{split}
\end{align}
which will have one more electron compared to the molecular doublet of interest.

For a molecular doublet with a single electron, we obtain the closed-shell mixed orbital reference
\begin{equation}
    \ketR = 
    {a}^\dagger_{I\alpha}{a}^\dagger_{I\beta}\ket{\text{vac}},\label{eq:one-orbital-system-1}
\end{equation}
which, expressed in the original basis,
\begin{equation}
\begin{split}
    \ketR = 
    \Big(&\cos^2\theta\;{a}^\dagger_{\mathcal{S}\alpha}{a}^\dagger_{\mathcal{S}\beta}
    + \sin^2\theta\;{a}^\dagger_{\mathcal{B} \alpha}{a}^\dagger_{\mathcal{B}\beta}\\
    - &\cos\theta\sin\theta(
    {a}^\dagger_{\mathcal{S} \alpha}{a}^\dagger_{\mathcal{B}\beta} 
    -{a}^\dagger_{\mathcal{S}\beta}{a}^\dagger_{\mathcal{B}\alpha})\Big)\ket{\text{vac}},\label{eq:one-orbital-system-2}
\end{split}
\end{equation}
{is seen to be the linear combination of the three singlet states obtained by placing two electrons in the two orbitals $\psi_\mathcal{S}$ and $\psi_\mathcal{B}$.}
In order to recover the determinant with a single electron of $\alpha$ spin in $\psi_\mathcal{S}$, we may act on $\ketR$ with the projection operator 
\begin{align}
    \mathcal{P} = n_{\mathcal{B}}^{\beta}(1-n_{\mathcal{B}}^{\alpha}) = P^\beta(1-P^\alpha) = P^\beta Q^\alpha,
\end{align}
where $P^\beta = n_{\mathcal{B}}^\sigma = E^\sigma_{\mathcal{B}\mathcal{B}}$ counts the number of
$\sigma$ electrons in $\psi_{\mathcal{B}}$:
\begin{align}
    P^{\beta}Q^{\alpha}\ketR
    &=
    {a}^\dagger_{\mathcal{S} \alpha}{a}^\dagger_{\mathcal{B} \beta}\ket{\text{vac}}( - \cos\theta\sin\theta).\label{eq:doublet-spin-projector}
\end{align}
The projection operator satisfies
\begin{align}
    \P^\dagger = \P,\quad
    \P ^2 = \P,
\label{eq:projection_prop}
\end{align}
and commutes with the Hamiltonian in eq \eqref{eq:hamiltonian}.

We define the entanglement coupled cluster wave function
\begin{align}
   \ket{\mathrm{ECC}} = \P \exp(T)\ketR, 
\end{align}
where the cluster operator $T$ is defined as in spin-adapted closed-shell coupled cluster theory\citep{helgaker2014} with respect to the mixed orbital basis. 

In the mixed orbital basis, the Hamiltonian becomes
\begin{align}
    H=\sum_{pq} \tilde{h}_{pq} E_{pq} + \frac{1}{2}\sum_{pqrs}\tilde{g}_{pqrs}e_{pqrs},\label{eq:hamiltonian-2}
\end{align}
where $\tilde{h}_{pq}$ and $\tilde{g}_{pqrs}$ are transformed to the mixed orbital basis, and the summations are over all orbitals. 
The number operator, used to define $\P$, is
\begin{align}
\begin{split}
    n_{\mathcal{B}}^{\sigma} =&\phantom{-}\sin^2\theta E_{II}^{\sigma} + \cos^2\theta E_{AA}^{\sigma}
    - \cos\theta\sin\theta(E^{\sigma}_{IA} + E^{\sigma}_{AI}),
\end{split}
\label{eq:number_op_1}
\end{align}
in the mixed orbital basis. 

By multiplying the Schrödinger equation from the left with $\exp(-T)$ and projecting onto the vectors $\{\braR, \bra{\mu}\}$, we obtain the equations for the ground state energy and amplitudes{:}
\begin{align}
    \Tbraket{\mathrm{R}}{\simP\simH}{\mathrm{R}} &= E_0 \Tbraket{\mathrm{R}}{\simP}{\mathrm{R}}\label{eq:ECC-omega}\\
    \Omega_\mu &= {\Omega}^S_\mu E_0, \label{eq:ECC-E}
\end{align}
where 
\begin{align}
    \Omega_\mu  &= \Tbraket{\mathrm{\mu}}{\simP\simH}{\mathrm{R}}\\
    \Omega^S_\mu &= \Tbraket{\mu}{\simP}{\mathrm{R}}.
\end{align}
Here, we have used $[H,\P] = 0$, and introduced the notation $\bar X = \exp(-T)X \exp(T)$.
The equations can be viewed as a change of the projection manifold compared to the standard closed-shell coupled cluster equations.

Entanglement coupled cluster excited states can be obtained within the equation-of-motion framework. We obtain the generalized eigenvalue equations
\begin{align}
    \boldsymbol{\simH R}^k = E_k\boldsymbol{\simS R}^k\label{eq:right}\\
\end{align}
where 
\begin{align}
\boldsymbol{\simH}=
    \begin{pmatrix}
        \Tbraket{\mathrm{R}}{\simP\simH}{\mathrm{R}} &\boldsymbol{{\eta}}^\mathrm{T}\\
        \boldsymbol{{\Omega}} &  \boldsymbol{J}
    \end{pmatrix},
\end{align}
and
\begin{align}
    \boldsymbol{\simS} =
        \begin{pmatrix}
        \Tbraket{\mathrm{R}}{\simP}{\mathrm{R}} &{\boldsymbol{\eta^S}}^\mathrm{T}\\
       \boldsymbol{\Omega^S}& \boldsymbol{{J^S}}
    \end{pmatrix},
\end{align} 
with
\begin{align}
\begin{split}
   \eta_\nu &=\Tbraket{\mathrm{R}}{\simP\simH}{\nu}\\
    J_{\mu\nu}&=\Tbraket{\mu}{\simP\simH}{\nu}\\
 {\eta}^S_{\nu} &= \Tbraket{\mathrm{R}}{\simP}{\mathrm{\nu}}.
\end{split}
\end{align}
The lowest generalized eigenvalue is the ground state energy and the remaining eigenvalues are excited state energies. 

To evaluate the ECC equations, we have first evaluated the projection manifold $\{\bra{\mu}\simP \}$.\cite{folkestad2023entanglement}
From eq. \eqref{eq:number_op_1}, we see that $\{\bra{\mu}\simP \}$ includes excited determinants of excitation rank $n + 2$ where $n$ is the order of the truncation of the cluster operator (i.e., the maximal excitation order of the vectors $\{\bra{\mu}\}$).  Therefore, in ECCSD, there are contributions to $\{\bra{\mu}\simP \}$  from singly, doubly triply, and quadruply excited determinants.

For the ground state, eqs. \eqref{eq:ECC-E} and \eqref{eq:ECC-omega}, a standard spin-adapted closed-shell coupled cluster implementation can be exploited to obtain all contributions to $\{\bra{\mu}\simP \}$ that are of excitation order 1 or 2. Terms arising from the reference contributions  to $\{\bra{\mu}\simP \}$  are trivial to implement.  Finally, the terms arising from projection of the Schrödinger equation on the triples and quadruples contributions  to $\{\bra{\mu}\simP \}$,
\begin{align}
    \bra{^{ab}_{ij}} E^\beta_{IA},\quad  \bra{^{a}_{i}} E^\alpha_{IA}E^\beta_{IA},\quad \bra{^{a}_{i}} E^\alpha_{jb}E^\beta_{IA},\quad \bra{^{ab}_{ij}} E^\beta_{IA}E^\alpha_{IA},
\end{align}
must be considered. We note that these contributions do not add terms which scale more steeply than $\mathcal{O}(N^6)$, since they maximally include two free (unrestricted) occupied and virtual indices.  These terms are implemented by use of automatic equation and code generation provided by the Julia package \textit{SpinAdaptedSecondQuantization\cite{sasq}} (SASQ).  The autogenerated terms have been compared to a previous proof-of-concept implementation, presented in Ref.~\citenum{folkestad2023entanglement}. 

The EOM equations are obtained by evaluating the linear transformation by the similarity transformed Hamiltonian
\begin{align}
    \rho_\eta = \Tbraket{\eta}{\simH}{\nu} c_\nu
\end{align}
and the linear transformation by the metric, 
\begin{align}
    \rho^S_\eta = \Dbraket{\eta}{\nu}c_\nu.
\end{align}
Here, the $\bra{\eta}$ vectors are those entering $\{\braR\simP, \bra{\mu}\simP\}$. 
The contributions to $\{\braR\simP, \bra{\mu}\simP\}$ that are singly and doubly excited determinants, can be partially extracted from a spin-adapted closed-shell adaptation of EOM coupled cluster. However, in this implementation all equations have been implemented using the automatic code generator.

\section{Results and Discussion}
{The ECCSD ground and excited states have been implemented in a development version of the e$^T$ program.\cite{folkestad2020t} The default convergence thresholds of e$^T$ v1.9.x have been used throughout: the Hartree--Fock equations are solved with a residual threshold (maximum norm) of $10^{-7}\;$a.u., Cholesky decomposition of the electron repulsion integrals is performed with a threshold of $10^{-4}\;$a.u., and the coupled cluster ground and excited state equations are solved to residual thresholds ($l_2$-norm) of $10^{-5}\;$a.u. and $10^{-3}\;$a.u., respectively. FCI calculations are solved to a residual threshold of $10^{-4}\;$a.u.}

\subsection{Comparisons to FCI for the water cation}
\begin{table*}[]
    \centering
    \begin{tabular}{c c c c c c c c}
    \toprule
    State & FCI $[\si{\hartree}]$  & ECCSD $[\si{\hartree}]$ & Error $[\si{\milli\hartree}]$ & Char. & $|c_{S,\text{max}}|^2$ & $|c_{D,\text{max}}|^2$ & $\frac{|\boldsymbol{R}_1|}{|\boldsymbol{R}|}$\\
    \midrule
     1    & -75.8069 & -75.8043  & 2.6 & --  & -- & --  & --\\
     2    & -75.7329 & -75.7302  & 2.7 & S   & $ 0.95 $ & $0.001$ & 0.99\\
     3    & -75.5582 & -75.5554  & 2.8 & S   & $ 0.94 $ & $0.002$& 0.99 \\
     4    & -75.2872 & -75.2743  & 12.9 & SD & $ 0.59 $ & $ 0.19 $ & 0.97\\
     5    & -75.2803 & -75.2780  & 2.3 & S   & $ 0.83 $ & $ 0.04 $ & 0.97 \\ 
     6    & -75.2290 & -75.1451 & 83.9 & SD  & $ 0.47 $ & $0.40$ & 0.37 \\ %
     7    & -75.2168 & -75.1988 & 18.0 & SD  & $ 0.56 $ & $0.23$ & 0.97\\
     8    & -75.1963 & -75.1923 & 4.0 & S    & $ 0.81 $ & $0.04$& 0.97\\ 
    \bottomrule
    \end{tabular}
    \caption{Total energies of the first eight states of H$_2$O$^+$ calculated with FCI and ECCSD. Energies are in $\si{\hartree}$ and energy differences in $\si{\milli\hartree}$. The ECCSD states are assigned to the FCI states based on a comparison of the ECCSD excitation vectors and the dominant configurations in the FCI state. To evaluate the character of the states, we consider the weight of the most important singly and doubly excited determinants contributing to the FCI state, $|c_{S,\text{max}}|^2$ and $|c_{D,\text{max}}|^2$. We can therefore characterize the FCI excited states according to wether they are dominated by single excitations of the reference (S) or they have significant doubles contributions (SD) (with $|c_{D,max}|^2 > 0.1$). Finally, we present the ECCSD single excitation vector contribution $\frac{|\boldsymbol{R}_1|}{|\boldsymbol{R}|}$.}
    \label{tab:fci_comparison}
\end{table*}
In Table \ref{tab:fci_comparison}, we compare ECCSD and FCI energies for the ground- and first excited states of H$_2$O$^+$ cation, using the cc-pVDZ basis. The ECCSD energies are all within $\SI{0.1}{\hartree}$ of the FCI energies, and for the ground state and some of the excited states, the error is of order $\SI{1}{\milli\hartree}$ 
Using the ECCSD code, we can also obtain the EMP2 energy, which for H$_2$O$^+$/cc-pVDZ is $\SI{-75.7096}{\hartree}$, giving an error of $\SI{97.2}{\milli\hartree}$ compared to FCI. 

We have characterized the FCI excited state vectors according to wether they are dominated by singly excited determinants (S), or wether they also have significant contributions from doubly exited determinants (SD), compared to the ROHF reference determinant with a single $\alpha$-electron in the SOMO. 
In particular, configurations with a single $\beta$-electron in the SOMO are generated by a double excitation of the reference, because it requires an excitation into the vacant mixed orbital, $\phi_A$, and out of the occupied mixed orbital, $\phi_I$. These \textit{spin-flip}  configurations (not to be confused with the spin-flip EOM method\citep{krylov2001size}) are the ones with significant contributions to the excited FCI vectors with SD character.   
The imbalance in the description of these states is due to the lack of spin completeness (described in Ref.~\citenum{herrmann2020generation}) i.e., that for a given spatial configuration, certain spin configurations are missing. To improve on this, triple excitations involving the mixed orbitals of the form 
$E_{bj}E_{Ai}E_{aI}$,
could be included in the cluster operator and (or only) in the EOM parametrization. Such excitations will not increase the overall scaling of the model (only the prefactors), but will, naturally, increase the complexity of the equations. 

We have assigned the ECCSD excited states to the FCI states by comparing the excitation vectors. The errors in the ECCSD energies are below $\SI{5}{\milli\hartree}$ for the states of S character, and at least an order of magnitude higher for the states with SD character. Except for state 6, all ECCSD excited states satisfy $\frac{|\boldsymbol{R}_1|}{|\boldsymbol{R}|}\geq0.97$. However, the ECCSD states that are assigned to the SD states all have significant contribution from spin-flip generating double excitations. The  ECCSD error increases with increasing weight of doubly excited determinants (measured by the square of the largest CI coefficient for a doubly excited determinant, $|c_{D,\text{max}}|^2$). 

\subsection{Satellite ionizations in Ethylene}

In Table \ref{tab:ethylene_adz}, we present the lowest excitations of the ethylene cation calculated with IP-EOM-CC3, IP-EOM-CCSD, and ECCSD using the aug-cc-pVDZ basis set. 
The calculated excited states are characterized according to whether they correspond to (direct) ionization of electrons in lower-lying valence orbitals or whether they are satellites (excited ionized states).  It is well established that the IP-EOM-CCSD approach cannot describe satellite states accurately, because these excitations effectively are pure double excitations of the closed-shell initial state. Hence, triple excitations in the parametrization (at least) is necessary to model these states through the IP-EOM approach. With ECCSD, satellites that are strongly dominated by single excitations of the cationic initial state, as is the case for the lowest satellite states of ethylene, can be modeled more accurately.
Similar results are found for the aug-cc-pVTZ basis set (see the \textit{Supporting Information}).

\begin{table*}[]
    \centering
    \begin{tabular}{c c c c c c}
    \toprule
     \multicolumn{2}{c}{IP-EOM-CC3}& \multicolumn{2}{c}{IP-EOM-CCSD} &  \multicolumn{2}{c}{EOM-ECCSD}\\
     \cmidrule(lr){1-2} \cmidrule(lr){3-4}\cmidrule(lr){5-6}
     $\omega [\si{\eV}]$ & Char. & $\omega [\si{\eV}]$ & Char. & $\omega [\si{\eV}]$ & Char. \\
     \midrule
      2.4811  & direct   & 2.4760  & direct   &  2.4808 & direct \\
      4.1419  & direct   & 4.2114  & direct   &  4.1879 & direct \\
      5.4670  & direct   & 5.5638  & direct   &  5.5699 & direct \\
      6.9547  & satellite & 8.9111  & direct   &  6.7310 & satellite \\
      8.1158  & satellite & 9.4264  & satellite &  8.0787 & satellite \\
      8.6710  & direct   & 10.7446 & satellite &  8.8686 & direct \\
      10.0948 & satellite & 13.2326 & satellite & 10.0136 & satellite  \\
         \bottomrule
    \end{tabular}
    \caption{Excitations of the ethylene cation, as calculated by IP-EOM-CC3, IP-EOM-CCSD and ECCSD, using the aug-cc-pVDZ basis set. The states are characterized according to whether they are satellites, i.e., an ionization accompanied by an excitation, or if they are (direct) ionizations from lower lying valence orbitals.}
    \label{tab:ethylene_adz}
\end{table*}
\subsection*{Core-excitations in the benzene cation}
\begin{table*}[]
\scriptsize
    \centering
    \begin{tabular}{c c c c c c c c}
    \toprule
      \multicolumn{4}{c}{Elongated} & \multicolumn{4}{c}{Contracted}\\
        \cmidrule(lr){1-4} \cmidrule(lr){5-8}
         CCSDT & CC3 & CCSD & ECCSD & CCSDT & CC3 & CCSD  & ECCSD   \\
        \midrule
     $283.22$ & $283.14$ ($0.08$)  &  $284.56$ ($1.34$) &  $284.22$ ($1.00$) & $283.21$ & $283.13$ ($0.08$) & $284.55$ ($1.34$) & $284.05$ ($0.84$) \\
     $283.23$ & $283.15$ ($0.08$)  &  $284.58$ ($1.35$) &  $284.22$ ($0.99$) & $283.21$ & $283.14$ ($0.07$) & $284.57$ ($1.36$) & $284.05$ ($0.84$) \\
     $283.24$ & $283.16$ ($0.08$)  &  $284.59$ ($1.35$) &  $284.25$ ($1.01$) & $283.25$ & $283.17$ ($0.08$) & $284.57$ ($1.32$) & $284.58$ ($1.33$) \\
     $283.26$ & $283.18$ ($0.08$)  &  $284.61$ ($1.35$) &  $284.25$ ($0.99$) & $283.26$ & $283.18$ ($0.08$) & $284.60$ ($1.34$) & $284.58$ ($1.32$) \\
     $283.29$ & $283.21$ ($0.08$)  &  $284.62$ ($1.33$) &  $284.70$ ($1.42$) & $283.28$ & $283.20$ ($0.08$) & $284.62$ ($1.34$) & $284.62$ ($1.34$) \\
     $283.29$ & $283.22$ ($0.06$)  &  $284.63$ ($1.34$) &  $284.71$ ($0.31$) & $283.29$ & $283.21$ ($0.08$) & $284.62$ ($1.33$) & $284.62$ ($1.33$) \\
     $289.06$ & $289.70$ ($0.64$)  &  $297.83$ ($8.77$) &  $288.75$ ($0.35$) & $289.01$ & $289.56$ ($0.55$) & $297.90$ ($8.89$) & $288.93$ ($0.08$) \\
     $289.10$ & $289.71$ ($0.61$)  &  $297.83$ ($8.73$) &  $288.75$ ($0.34$) & $289.01$ & $289.60$ ($0.59$) & $297.90$ ($8.89$) & $288.93$ ($0.08$) \\
     $289.11$ & $289.72$ ($0.61$)  &  $298.10$ ($8.99$) &  $289.45$ ($0.35$) & $289.02$ & $289.62$ ($0.60$) & $297.90$ ($8.88$) & $288.95$ ($0.07$) \\
     $289.11$ & $289.72$ ($0.61$)  &  $298.10$ ($8.99$) &  $289.46$ ($0.35$) & $289.02$ & $289.62$ ($0.60$) & $297.90$ ($8.88$) & $288.95$ ($0.07$) \\
     $289.11$ & $289.72$ ($0.61$)  &  $298.10$ ($8.99$) &  $289.46$ ($0.35$) & $289.29$ & $289.95$ ($0.66$) & $298.24$ ($8.95$) & $289.77$ ($0.48$) \\
     $289.53$ & $289.72$ ($0.19$)  &  $298.10$ ($8.57$) &  $289.46$ ($0.07$) & $289.29$ & $289.95$ ($0.66$) & $298.24$ ($8.95$) & $289.77$ ($0.48$) \\
         \bottomrule
     \end{tabular}
    \caption{Core excitations of the benzene cation calculated with CVS-IP-EOM-CCSDT (CCSDT), 
    CVS-IP-EOM-CC3 (CC3), CVS-IP-EOM-CCSD (CCSD), and CVS-EOM-ECCSD (ECCSD) using the cc-pVDZ basis set. All energies are given in $\si{\eV}$. Absolute errors of ECCSD compared to CCSDT are given in parenthesis.}
    \label{tab:benzene_core}
\end{table*}

Finally, we consider the benzene molecule, which undergoes symmetry breaking due to the Jahn-Teller effect upon ionization. The cation has two minima, with lowered symmetry ($D_{2h}$) compared to the neutral molecule ($D_{6h}$).\citep{vidal2020interplay} We use the two optimized geometries (elongated and compressed) from the \textit{Supporting Information} of Ref. \citenum{vidal2020interplay}.  We report 
the carbon K-edge excitation energies of the benzene cation for the two geometries.  The core excitations are calculated using the core-valence separation (CVS) approximation.\citep{coriani2015communication,coriani2016erratum} 
In the CVS approximation, core excitations are considered by neglecting any pure valence amplitudes in the excitation vectors. 
While the CVS approximation is implemented through projection of the full space excitation vectors for ECCSD, CCSD, and CCSDT, only the contributing terms are calculated for the CC3 model.\citep{paul2020new} The CVS-IP-EOM models are implemented by neglecting all elements of the excitation vectors that do not contain an occupied index corresponding to a core orbital and a virtual index corresponding to the non-interacting bath orbital. For these models, the core excitation energies of the cation are obtained by subtracting the valence IP from the core IPs.
We present calculations with CVS-IP-EOM-CCSD and CVS-IP-EOM-CC3, CVS-IP-EOM-CCSDT of the neutral molecule and CVS-EOM-ECCSD of the cation. The cc-pVDZ basis set is used, additional calculations with the aug-cc-pVDZ basis can be found in the \textit{Supporting Information}. 

From these calculations, we observe that, while CVS-IP-EOM-CCSD provides an adequate model of the first group of six core excitations, the second group of six excitations has an error of almost $\SI{10}{\eV}$ compared to the \textcolor{black}{CVS-IP-EOM-CCSDT} calculations. These constitute satellite states of the core-ionized molecule. The large errors arise because of the low degree of parametric freedom resulting from projecting out all excitations that do constitute a core ionization. We may compare these results to those obtained for satellites in the valence ionized ethylene molecule of Table \ref{tab:ethylene_adz}. 
From the CVS-EOM-ECCSD model, we expect a quality comparable to the CVS-EOM-CCSD model for core excited states (i.e., using the CVS projection, but not the IP projection). The core excitation energies of the benzene cation are found to have absolute errors well within $\SI{2}{\eV}$ compared to EOM-IP-CCSDT. 

\section{Conclusions}
We have presented the extension of the entanglement coupled cluster singles and doubles (ECCSD) model to the spin-adapted excited states of doublet systems, through the equation-of-motion (EOM) formulation.
For states that are dominated by single substitutions compared to the ROHF reference determinant, the accuracy is comparable to that of spin-adapted closed-shell CCSD. While the model is spin-adapted, the parametrization is not spin-complete. The addition of triple excitations in the parametrization that includes an excitation into and out of the mixed orbitals ($E_{bj}E_{Ai}E_{aI}$) could improve the accuracy for states dominated by configurations where the singly occupied orbital of the high-spin reference determinant is occupied by a single $\beta$-electron.

We have also demonstrated that the EOM-ECCSD method can be used in situations where IP-EOM-CCSD cannot be trusted. 

\section{Acknowledgments}
This work has received funding from the European Research Council (ERC) under the European Union’s Horizon 2020 Research and Innovation Programme (grant agreement No. 101020016).
\bibliography{main}
\end{document}